\documentclass[twocolumn,preprintnumbers,amsmath,amssymb]{revtex4}

\usepackage{graphicx}
\usepackage{dcolumn}
\usepackage{bm}

\begin{document}

\title{Magnetic and Electronic Properties of the New Ferrimagnet 
Sr$_8$CaRe$_3$Cu$_4$O$_{24}$}

\author{Masanori Kohno, Xiangang Wan and Xiao Hu}
\affiliation{Computational Materials Science Center, National Institute 
for Materials Science, Tsukuba 305-0047, Japan}

\date{\today}

\begin{abstract}
Magnetic and electronic properties of the recently-discovered material 
Sr$_8$CaRe$_3$Cu$_4$O$_{24}$ were investigated by means of a quantum Monte 
Carlo simulation, the Green function method and the LSDA+$U$ (local 
spin-density approximation plus the Hubbard-$U$ term) method. The LSDA+$U$ 
calculation shows that the ground state is an insulator with magnetic moment 
$M$=1.01$\mu_{\rm B}$/f.u., which is consistent with experimental results. 
The magnetic sites were specified and an effective model for the magnetic 
properties of this compound derived. The resultant effective model is a 
three-dimensional Heisenberg model with spin-alternation. Finite-temperature 
properties of this effective model are investigated by the quantum Monte 
Carlo method (continuous-time loop algorithm) and the Green function method. 
The numerical results are consistent with experimental results, indicating 
that the model is suitable for this material. Using the analysis 
of the effective model, some predictions for the material are made.
\end{abstract}

\maketitle

\section{Introduction}
\label{sec:intro}
Strong electronic correlations produce a rich variety of phenomena. 
High-$T_{\rm c}$ superconductivity\cite{high_Tc} is one of such a phenomenon. 
It is widely believed that the origin of high-$T_{\rm c}$ superconductivity 
is related to the nature of undoped systems\cite{Zhang_Rice}, which are Mott 
insulators with an antiferromagnetic order. Recently-discovered perovskite 
cuprate Sr$_8$CaRe$_3$Cu$_4$O$_{24}$\cite{exp} is a magnetic insulator which 
has spontaneous magnetization at room temperature and in some respects 
resembles parent materials of high-$T_{\rm c}$ superconductors. It is 
expected that the ferromagnetism of Sr$_8$CaRe$_3$Cu$_4$O$_{24}$ might be 
caused by strong electronic correlations due to a similar mechanism to the 
antiferromagnetism of parent materials of high-$T_{\rm c}$ superconductors. 
Furthermore, among ferromagnetic cuprates this material has an unusually high 
magnetic transition temperature ($T_{\rm c}$) up to 440K, although those of 
other ferromagnetic cuprates are at most 30K\cite{ferroCup}. Thus, the 
material is interesting for its high $T_{\rm c}$ induced by strong electronic 
correlations. The purpose of this paper is to clarify the origin of magnetism 
and to predict the possible magnetic properties of this material. 

\section{Local Spin-Density Approximation}
\label{sec:method}
\begin{figure}
\includegraphics[scale=0.20]{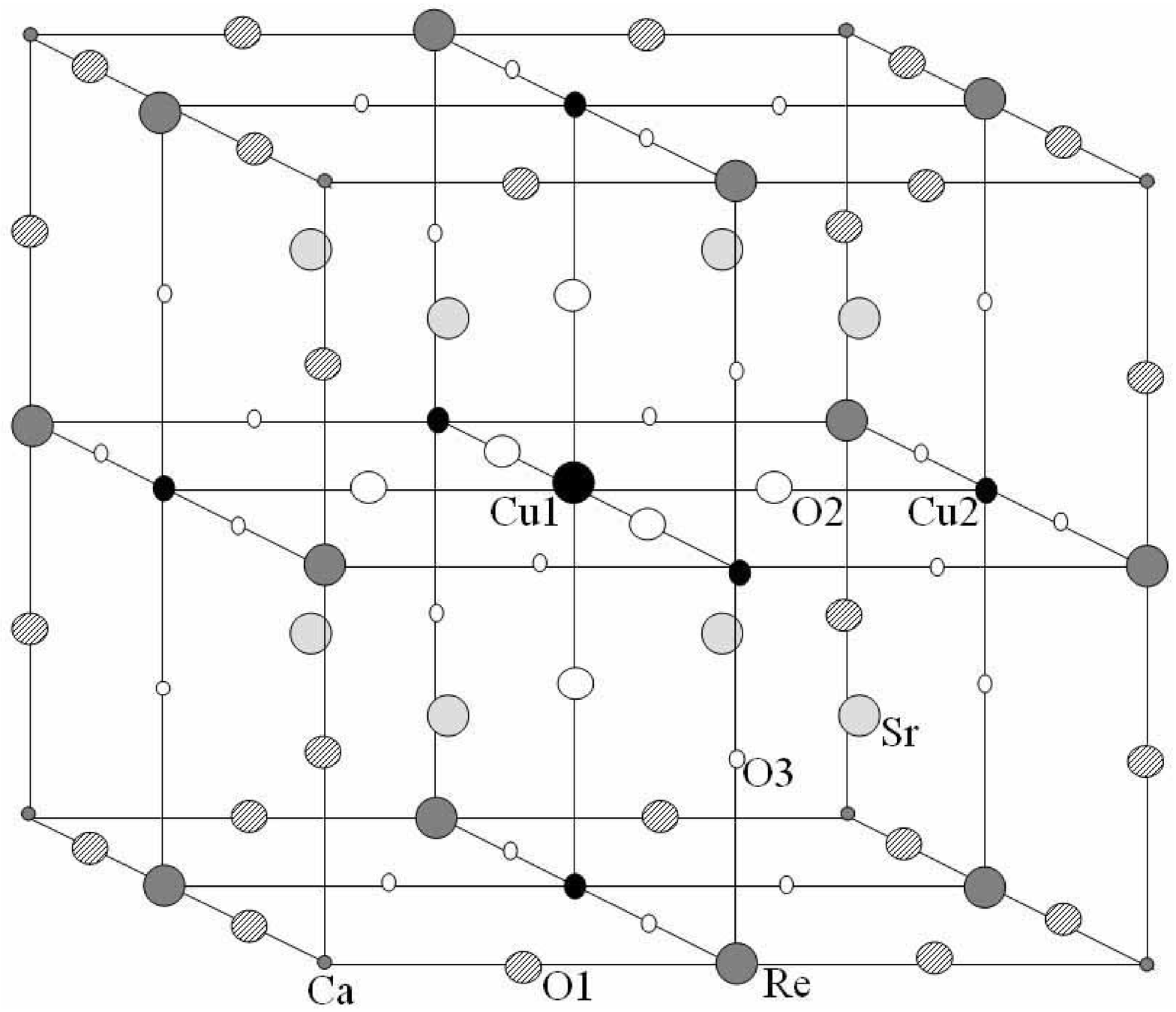}
\caption{Unit cell of Sr$_8$CaRe$_3$Cu$_4$O$_{24}$.}
\label{fig:lattice}
\end{figure}

The lattice structure is shown in Fig. \ref{fig:lattice}, which is determined 
experimentally\cite{exp}. 
In order to investigate electronic properties of the compound, the local 
spin-density approximation plus the Hubbard-$U$ term (LSDA+$U$) method using 
the package WIEN2K\cite{Wien2k} which is an implementation of the 
density-functional APW-lo method\cite{APW}, has been used. The cutoff values 
were set at $RK_{\rm max}$=7, the LSDA+$U$ parameters $U$=10eV and 
$J_{{\rm LSDA}+U}$=1.2eV\cite{J_LSDA}. For the exchange correlation, the 
standard generalized gradient approximation (GGA)\cite{GGA} has been applied. 
The numerical results show that the ground-state is an insulator with a 
ferrimagnetic order: The charge gap is estimated as about 1.68eV. Magnetic 
moments are almost localized at Cu sites, and their directions at Cu1 and Cu2 
sites are opposite. The absolute values of the magnetic moments at Cu1 and 
Cu2 are 1.01$\mu_{\rm B}$ and 0.86$\mu_{\rm B}$ per formula unit (f.u.), 
respectively, and those of other sites are less than 0.07$\mu_{\rm B}$/f.u.. 
The total magnetic moment is 1.01$\mu_{\rm B}$/f.u., which is comparable to 
the experimental result (0.95$\mu_{\rm B}$/f.u.)\cite{exp}. The LSDA+$U$ 
calculation shows that the orbital degrees of freedom are also ordered. The 
high $T_{\rm c}$ would be due to the large overlap between orbitals of O2 and 
Cu originated from orbital ordering. The mechanism of the antiferromagnetic 
coupling is the super-exchange\cite{superex} as in the parent materials of 
high-$T_{\rm c}$ superconductors. Details of the LSDA+$U$ calculation are 
presented in a separate paper\cite{Wan}.
\section{Effective model for magnetism}
\label{sec:model}
\begin{figure}
\includegraphics[scale=0.20]{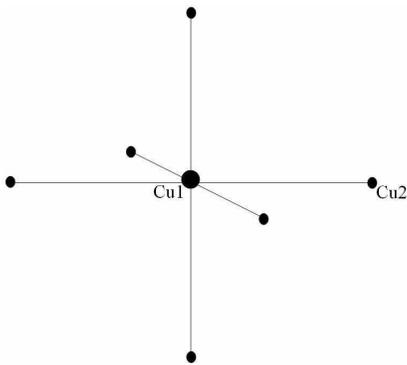}
\caption{Unit cell of the effective model.}
\label{fig:model}
\end{figure}

By using LSDA+$U$ method, the magnetic sites of this compound have been 
identified. Since the system is an insulator, it is natural to expect that 
the effective model for magnetism is a Heisenberg model. It was assumed that 
the spins on nearest-neighbor sites interact with each other. Then, the 
Hamiltonian of the effective model is expressed as
\begin{equation}
\label{eq:model}
{\cal H}=J\sum_{i,p}\mbox{\boldmath $S$}_i\cdot
\mbox{\boldmath $s$}_{i+\frac{p}{2}},
\end{equation}
where ${\mbox{\boldmath $S$}}_i$ and ${\mbox{\boldmath $s$}}_{i+p/2}$ denote 
spin operators at Cu1 ($i$) and Cu2 ($i$+$p/2$) sites, respectively. Here, 
$p$ represents the unit vectors ($p$=$\pm{\hat x}$, $\pm{\hat y}$, 
$\pm{\hat z}$). The unit cell of this effective model is illustrated in 
Fig. \ref{fig:model}.
\par
Consider the appropriate combination of spin length ($S$, $s$), where $S$ and 
$s$ denote the spin length at Cu1 and Cu2 sites, respectively. In order to 
determine these, the magnetization of the ground state ($M_{\rm GS}$) and 
that of the fully-polarized state ($M_{\rm FP}$) for the effective model 
in a unit cell were calculated. In the case of ($S$, $s$)=(1/2, 1/2), 
$M_{\rm GS}$=1 and $M_{\rm FP}$=2, which correspond to 2$\mu_{\rm B}$/f.u. 
and 4$\mu_{\rm B}$/f.u., by setting the $g$-factor 2. For ($S$, $s$)=
(1/2, 1), (1, 1/2), (1, 1), ($M_{\rm GS}$, $M_{\rm FP}$) becomes (5/2, 7/2), 
(1/2, 5/2), (2, 4), respectively. Here, $M_{\rm GS}$ and $M_{\rm FP}$ 
obtained by the LSDA+$U$ method are 1.01$\mu_{\rm B}$/f.u. and 
5$\mu_{\rm B}$/f.u.. Hence, ($S$, $s$)=(1, 1/2) is the most suitable for 
this compound. 
\par
It should be noted that $M_{\rm GS}$ and $M_{\rm FP}$ of the effective model 
are exactly 1/2 and 5/2 (equal to 1$\mu_{\rm B}$/f.u. and 
5$\mu_{\rm B}$/f.u.) in any size of systems due to the 
Marshall-Lieb-Mattis theorem\cite{MLM_theorem}. The ground-state 
magnetization of the effective model (1$\mu_{\rm B}$/f.u.) is comparable 
to the experimental results\cite{exp}. 
\par
Next, the strength of antiferromagnetic coupling $J$ was considered. The 
energy difference between the ground state and the fully-polarized state 
in a unit cell was calculated by the exact diagonalization method. The result 
is 8$J$. The energy difference was calculated for Sr$_8$CaRe$_3$Cu$_4$O$_{24}$ 
also by the LSDA+$U$ method. The result was 0.036{\it Ry}. Hence, by 
comparing these results, $J$=0.0045{\it Ry}(=710K) was obtained. 
\section{Green function method}
\label{sec:Greenfunc}
Using the above arguments, the effective model for magnetism of 
Sr$_8$CaRe$_3$Cu$_4$O$_{24}$, which is a spin-alternating Heisenberg model in 
three-dimensions has been obtained. In order to investigate 
finite-temperature properties of this compound, the Green function 
method\cite{GFmethod} was applied to the effective model where Green 
functions are defined as Fourier components of time-dependent correlation 
functions:
\begin{equation}
G^{Ss}(k,\omega)=\frac{1}{N}\int dt\sum_{i,j}\langle\langle S_i^+(t);
s_{j+\frac{p}{2}}^-\rangle\rangle{\rm e}^{{\rm i}k
\left(r_i-r_{j+\frac{p}{2}}\right)-{\rm i}\omega t},
\label{eq:g}
\end{equation}
where $N$ is the number of unit cells. Since there are four sites in a 
unit cell, 16(=4$\times$4) Green functions are necessary. Here, the 
double-bracket correlation function is defined as 
\begin{equation}
\langle\langle A(t);B\rangle\rangle\equiv-\theta(t)\langle[A(t),B]\rangle. 
\end{equation}
Coupled equations were obtained for 16 Green functions by using a decoupling 
approximation\cite{decouple}
\begin{equation}
\langle\langle S^z(t)s^+(t);s^-\rangle\rangle\simeq
\langle S^z\rangle\langle s^+(t);s^-\rangle. 
\end{equation}
These equations were solved analytically and the temperature dependence of 
magnetizations was obtained using these Green functions by following Callen's 
scheme\cite{Callen}. 
\begin{figure}
\includegraphics[scale=0.13]{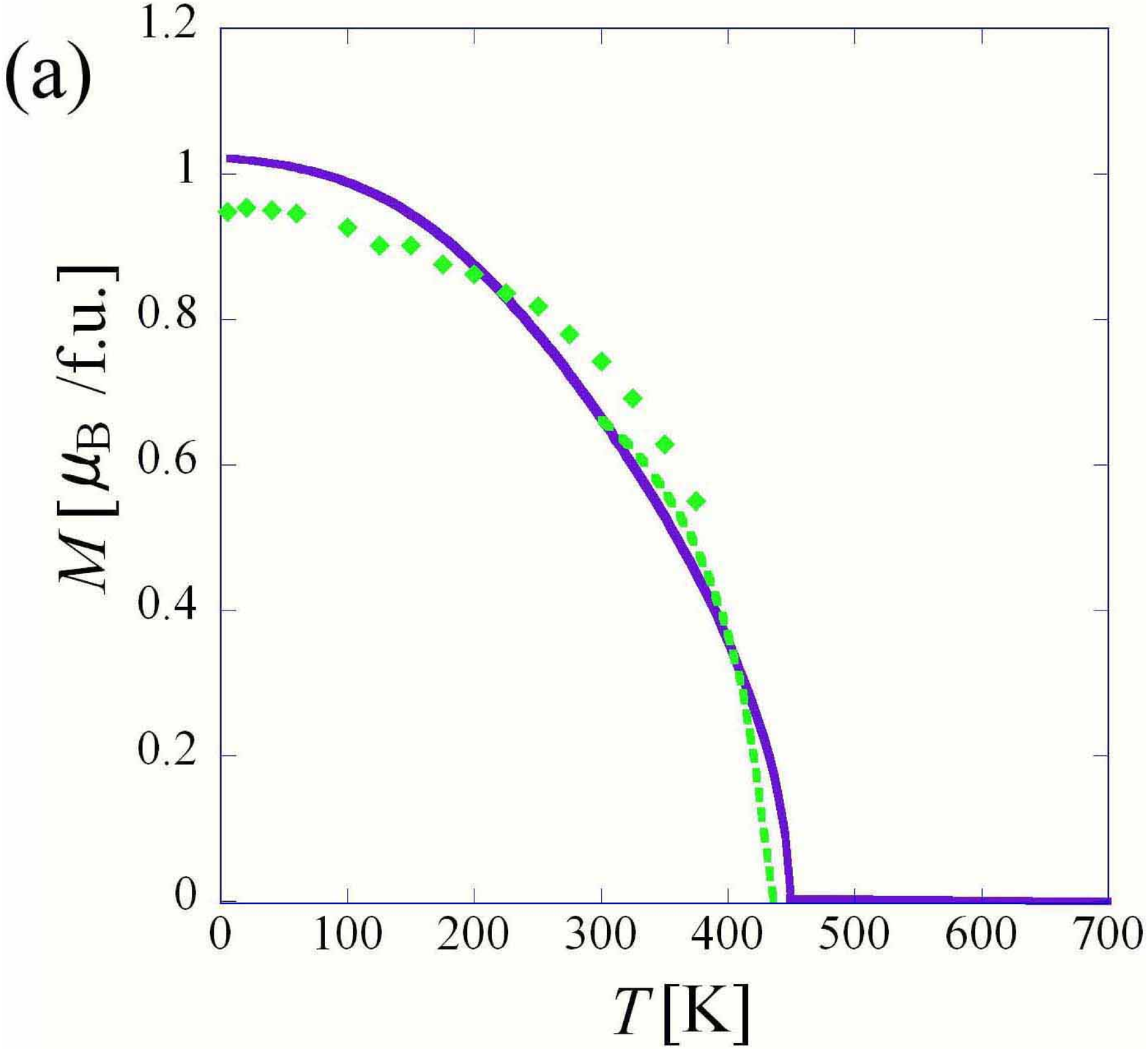}\\
\vspace{10pt}
\includegraphics[scale=0.13]{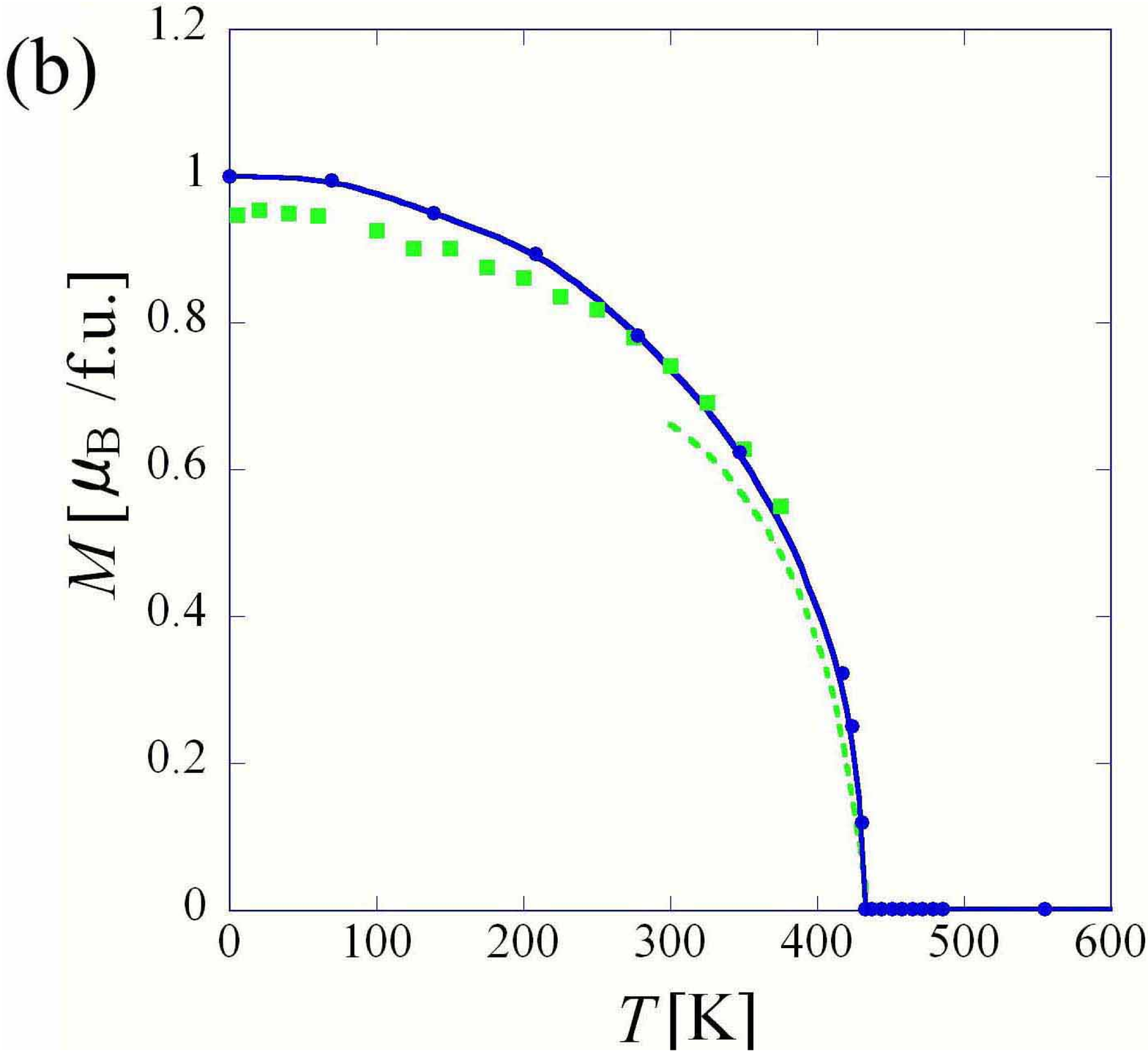}
\caption{Temperature dependence of spontaneous magnetization obtained by 
(a) the Green function method and (b) the quantum Monte Carlo method 
(solid curves). Dots and dotted line denote the experimental 
results\cite{exp}.}
\label{fig:MT}
\end{figure}
Figure \ref{fig:MT} (a) shows the result of the Green function method. 
The agreement with the experimental result is good, indicating that the 
effective model is suitable for describing the magnetic properties of 
Sr$_8$CaRe$_3$Cu$_4$O$_{24}$. From the LSDA+$U$ result, $J$ was set as 
$J$=710K. 
\section{Quantum Monte Carlo Simulation}
\label{sec:QMC}
In order to investigate magnetic properties of this compound more 
quantitatively, a quantum Monte Carlo method was applied to the 
effective model. The algorithm used was the continuous-time loop 
algorithm (see Refs. in \cite{loopalg}), which is a powerful method for 
non-frustrated spin systems. More than one million updates for each 
simulation were performed. The system size was up to 16$\times$16$\times$16 
unit cells which correspond to 16,384 sites. 
\par
Spontaneous magnetization can be obtained by extrapolating ${\bar S}(N)$ 
which is defined as ${\bar S}(N,T)\equiv\sqrt{3\langle S_0^zS_l^z\rangle_T}$, 
where $S_0^z$ and $S_l^z$ are the $z$-component of Cu1-spin at the center of 
the system and that of the furthest site from the center, respectively. Here, 
$N$ denotes the number of unit cells. By extrapolating ${\bar S}(N,T)$ in the 
thermodynamic limit, we obtain the temperature dependence of spontaneous 
magnetization as shown in Fig. \ref{fig:MT} (b). The quantum Monte Carlo 
result is consistent with experimental results. By tuning $J$ to fit the 
transition temperature, we obtain $J$=695K. The first estimation of $J$=710K 
based on LSDA+$U$ results was not far from the fitted value, suggesting that 
the above mapping is valid. To date, the temperature dependence of the 
spontaneous magnetization is the only quantity that has been measured 
experimentally\cite{exp}. However, since the effective model for this 
compound has been derived, further predictions of its magnetic properties 
through the investigation of this effective model can be made. 
\par
\begin{figure}
\includegraphics[scale=0.13]{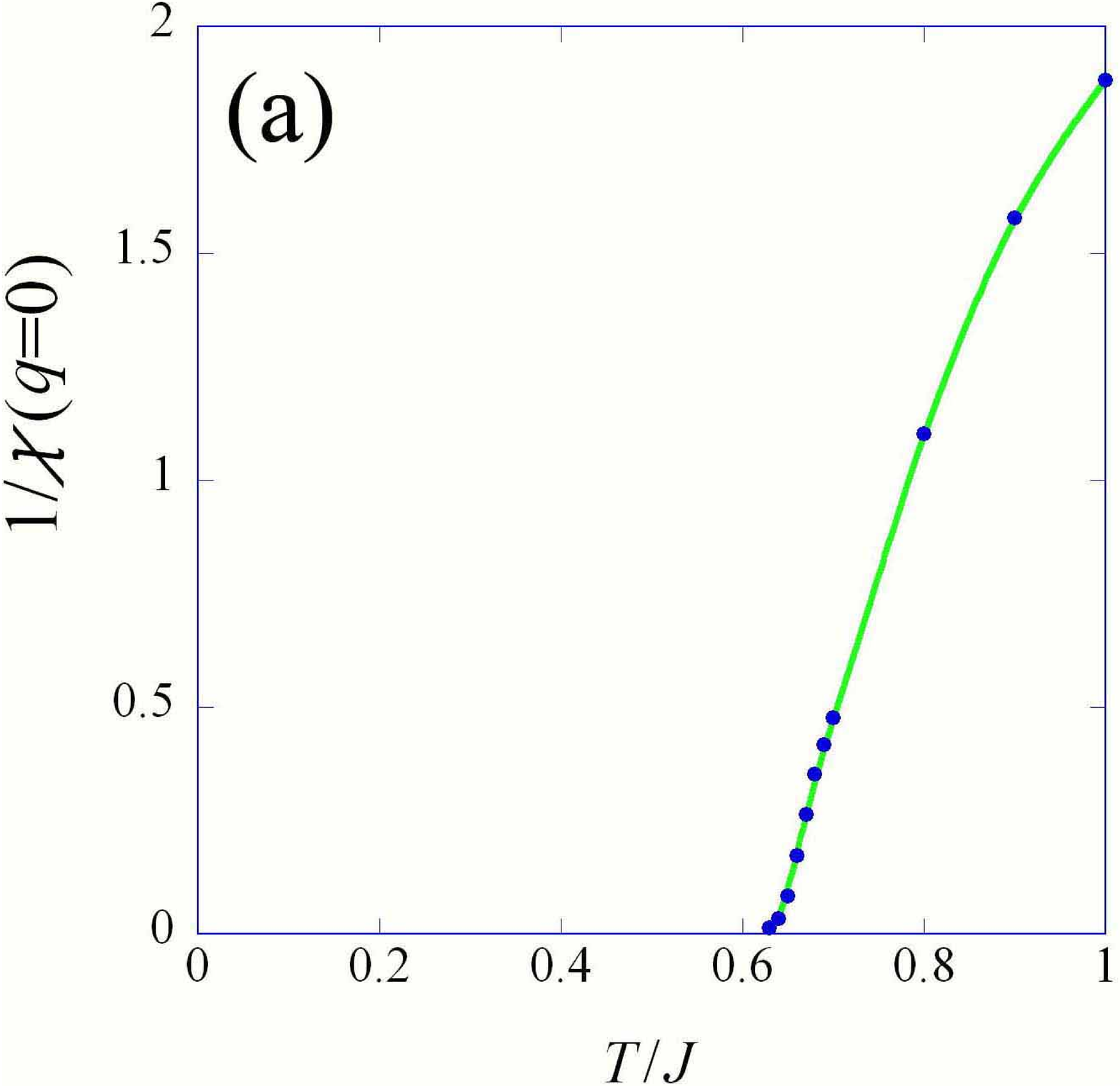}\\
\vspace{10pt}
\includegraphics[scale=0.13]{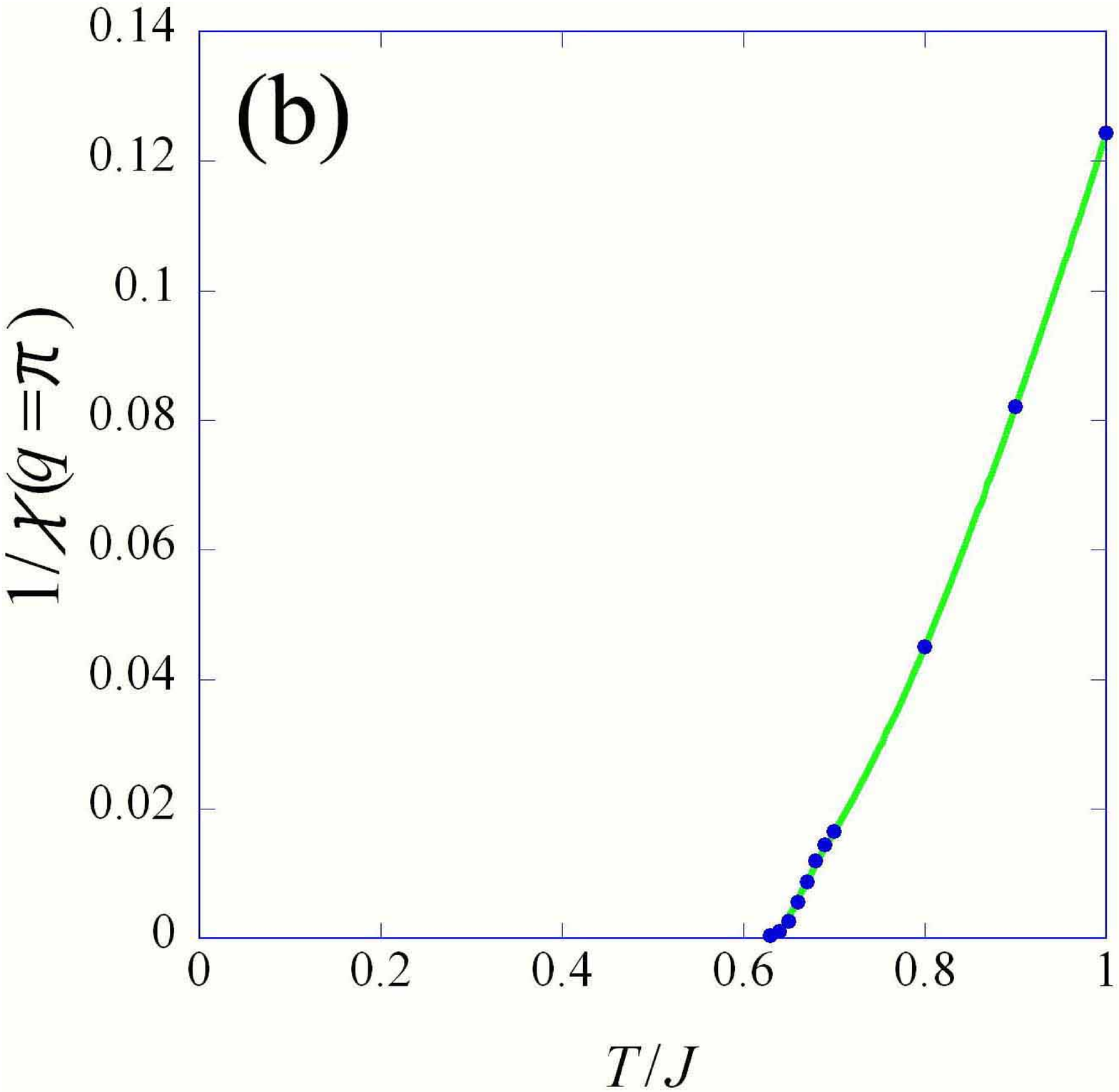}
\caption{Temperature dependence of the inverse of uniform (a) and 
staggered susceptibilities (b) above the critical temperature.}
\label{fig:sus}
\end{figure}
\begin{figure}
\includegraphics[scale=0.13]{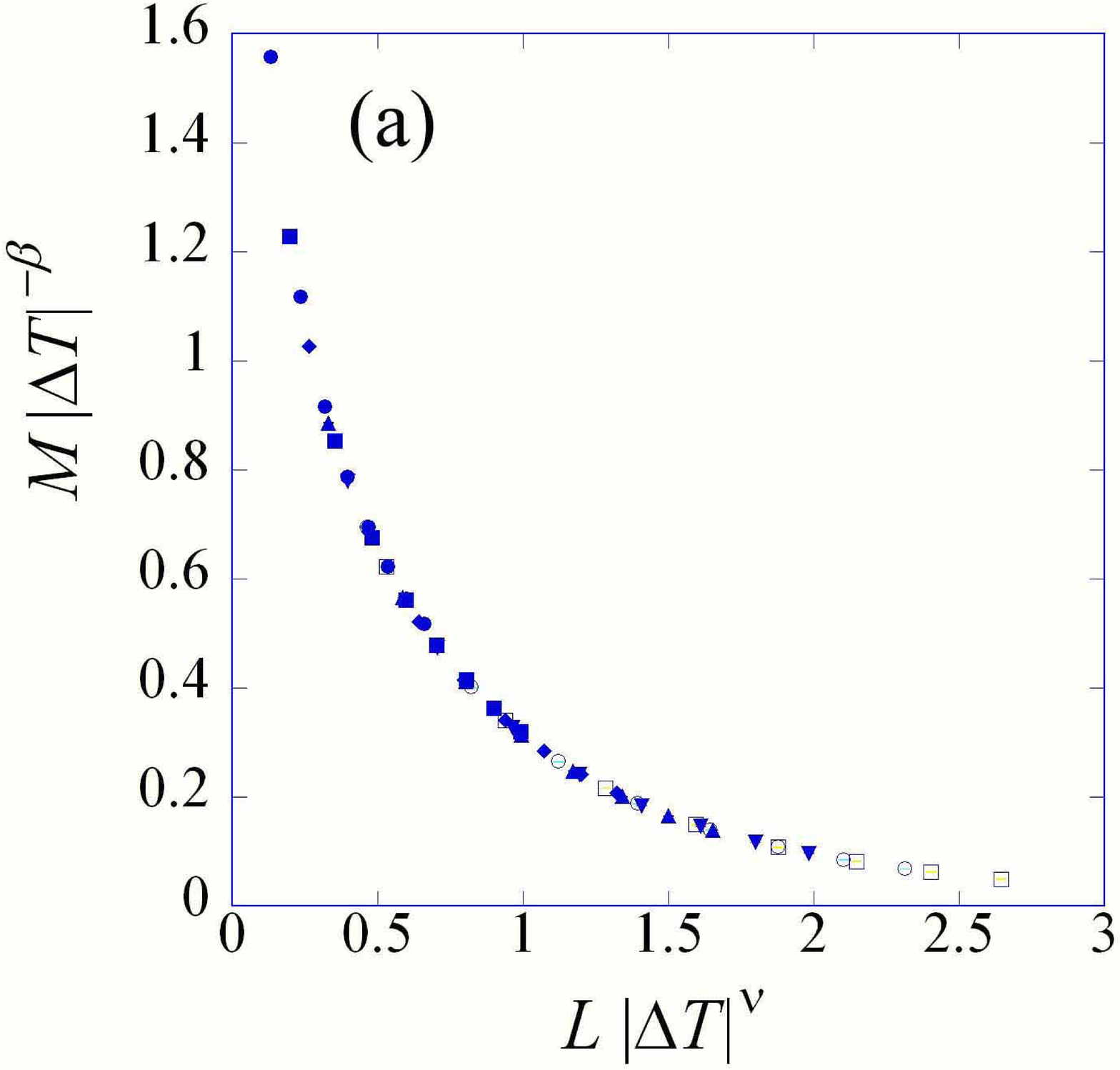}\\
\vspace{10pt}
\includegraphics[scale=0.13]{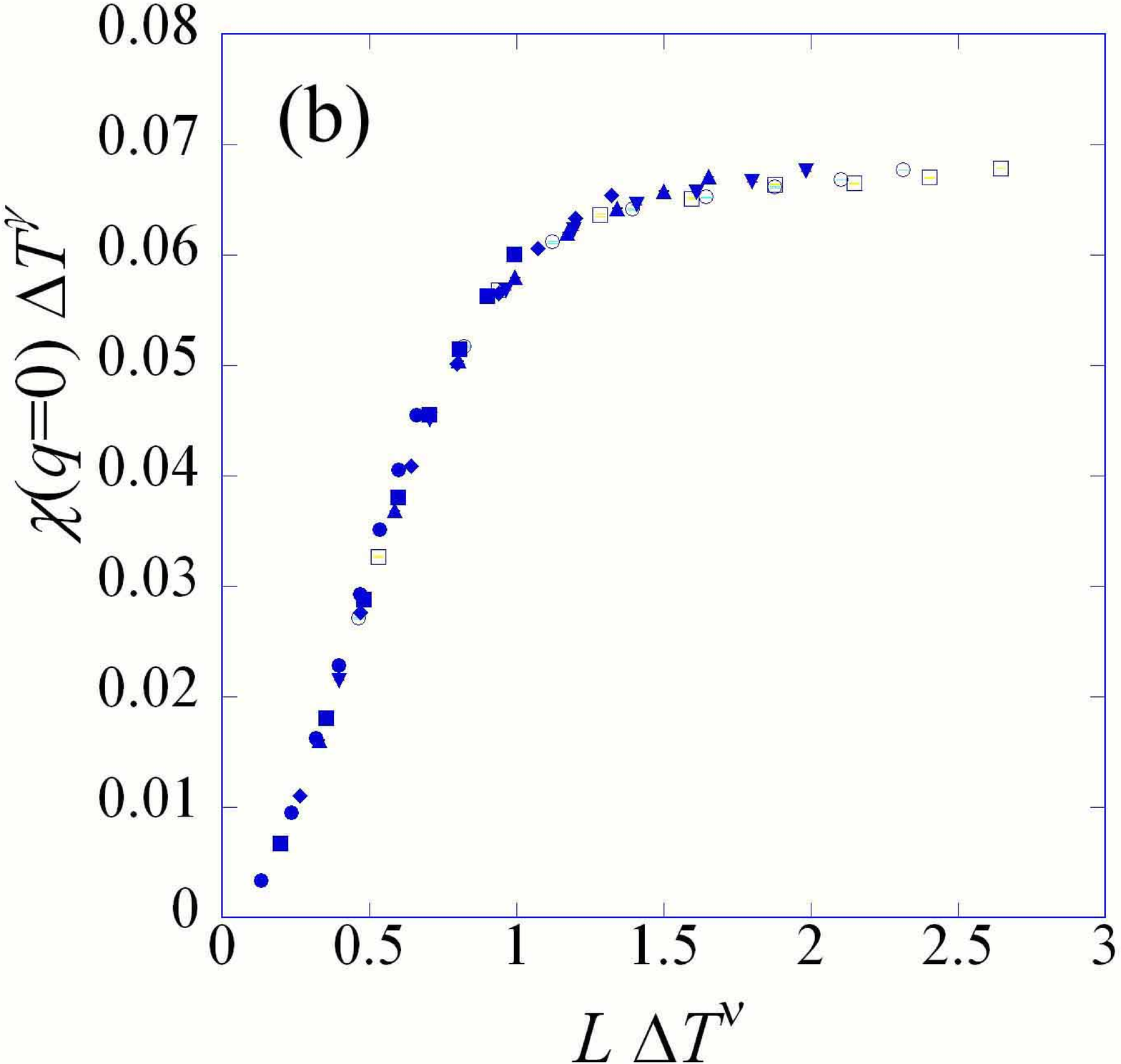}\\
\vspace{10pt}
\includegraphics[scale=0.13]{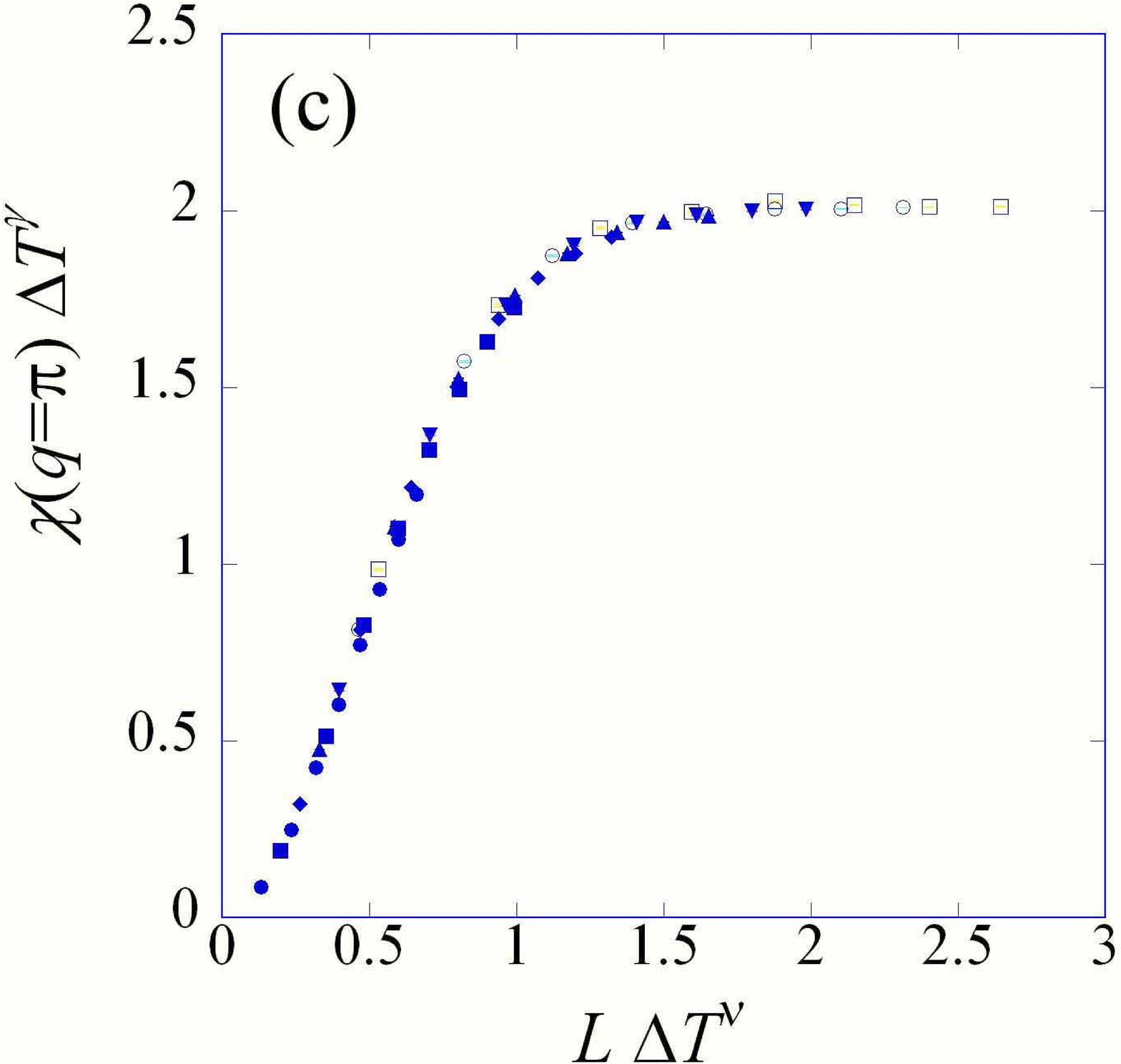}
\caption{Scaling plot for $M$ and $\chi$. We set the exponents of the 
three-dimensional (3D) Heisenberg model: $\nu$=0.7054, $\beta$=0.3646 
and $\gamma$=1.3866\cite{3DHeis}. The data are obtained in clusters of 
$L$=4$\sim$16 in the temperature range of $T$/$J$=0.63$\sim$0.7.}
\label{fig:scaling}
\end{figure}

The temperature dependence of the inverse of uniform and staggered 
susceptibilities above the critical temperature is shown in 
Fig. \ref{fig:sus}. 
As is evident from this figure, both susceptibilities diverge toward the 
critical point. In general, physical quantities show a power-law behavior 
at the critical point of the second-order phase transition. The universality 
class is characterized by the exponents of physical quantities such as 
length scale $\xi\propto\Delta T^{-\nu}$, 
magnetization $M\propto|\Delta T|^{\beta}$  and susceptibility 
$\chi\propto\Delta T^{-\gamma}$, where $\Delta T\equiv T-T_{\rm c}$. 
In finite-size systems, the magnetization and the susceptibility can be 
expressed in terms of scaling functions as\cite{scalingplot}
\begin{equation}
M\simeq|\Delta T|^{\beta}{\bar M}(L|\Delta T|^{\nu}),
\end{equation}
\begin{equation}
\chi\simeq\Delta T^{-\gamma}{\bar \chi}(L\Delta T^{\nu}),
\end{equation}
where $L$ is the number of unit cells in the $x$-, $y$- or $z$-direction. 
$M|\Delta T|^{-\beta}$ and $\chi\Delta T^{\gamma}$ were plotted as a function 
of $L|\Delta T|^{\nu}$ in Fig. \ref{fig:scaling}, where we use the exponents 
of the three-dimensional (3D) Heisenberg model: $\nu$=0.7054, $\beta$=0.3646 
and $\gamma$=1.3866\cite{3DHeis}. The figure shows that data in various sizes 
and various temperatures near the critical point fall into single curves, 
suggesting that the universality class is that of the 3D Heisenberg model. 
\par
\begin{figure}
\includegraphics[scale=0.13]{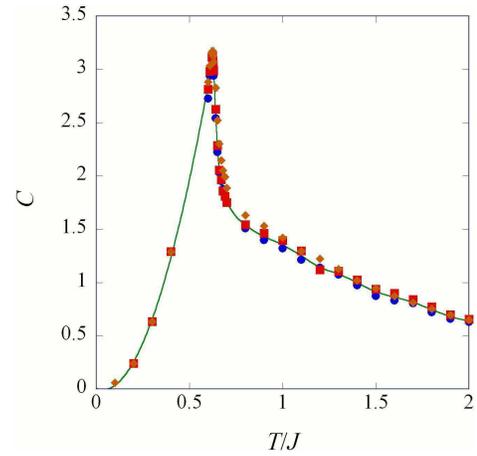}
\caption{Temperature dependence of the specific heat. Circles, squares and 
diamonds denote the data of $L$=12, 14 and 16, respectively. Solid line is 
a guide to the eye.}
\label{fig:CT}
\end{figure}

The temperature dependence of the specific heat is plotted in 
Fig. \ref{fig:CT}. 
The size dependence of the specific heat is small. The maximum value is 
almost saturated at about 3.1. The critical temperature was estimated as 
0.622$\pm$0.002, assuming that the specific heat is a maximum at the 
critical temperature. Since the specific heat has little size-effect, 
it is expected that the specific heat at the critical temperature remains 
finite in the thermodynamic limit. This is consistent with the scaling 
property of the 3D Heisenberg model: $C\propto\Delta T^{-\alpha}$, 
$\alpha$=$-$0.1162$<$0\cite{3DHeis}.
\section{Conclusions}
\label{sec:conclusions}
The magnetic and electronic properties of Sr$_8$CaRe$_3$Cu$_4$O$_{24}$ have 
been investigated by means of the quantum Monte Carlo method, 
the Green function method and the LSDA+$U$ method. Since the LSDA+$U$ 
calculation shows that magnetic moments are localized at Cu sites, a 
Heisenberg model was introduced as an effective model for magnetic properties. 
The spin lengths at Cu1 and Cu2 sites as one and one half, respectively, were 
obtained by comparing the result given by the effective model in a unit cell 
and that given by the LSDA+$U$ method. The finite-temperature properties of 
the effective model were investigated by the Green function method and the 
quantum Monte Carlo method. The agreement on magnetization at finite 
temperatures between the numerical results and the experimental results was 
good, suggesting that the effective model is suitable for describing the 
magnetic properties of Sr$_8$CaRe$_3$Cu$_4$O$_{24}$. The properties of the 
effective model were further investigated and revealed that the universality 
class of the critical point is that of the 3D Heisenberg model. Some of the 
results given by the effective model may be accessible by experiments. 
\section*{Acknowledgements}
Thanks are expressed to Dr. E. Takayama-Muromachi for drawing attention to 
this material and for valuable discussions. Dr. C. Yasuda, Dr. M. Arai and 
Dr. M. Katakiri are appreciated for technical help and thanks are due to 
Dr. A. Tanaka for discussions. This work was partially supported by Japan 
Society for the Promotion of Science (Grant-in-Aid for Scientific 
Research (C) No. 15540355).

\end{document}